\documentclass[11pt,twoside]{article} 
\usepackage{acta-info}
\usepackage{euler}

\newcommand{\vol}{\textbf{3,} 2 (2011) 141--157}   %% to be completed by the editor, do not modify it
\newcommand{\amss}{\amssubj{%
68N19
}}     %% Source:   http://www.ams.org/msc/
\newcommand{\CRs}{\CRsubj{%
D.3.2
}}   %% Source: http://www.acm.org/about/class/1998
\newcommand{\keyw}{\keywords{%
C++, STL, template, compilation
}}

\begin{document}

\title{% 
%%Write here the title of your paper, as
C++ Standard Template Library by template specialized containers
}
\maketitle
%% SINGLE AUTHOR. If you are a single author, please, use the following command and delete
%%                the \twoauthors command completely.

\oneauthor{
Norbert PATAKI
}{
\href{http://www.inf.elte.hu/english/aboutus/departments/Lapok/DepartmentofProgrammingLanguagesAndCompilers.aspx}{Dept. of Programming Languages and Compilers}\\
\href{http://www.inf.elte.hu/english/Lapok/default.aspx}{Faculty of Informatics}, \href{http://www.elte.hu/en}{E\"otv\"os Lor\'and University}\\ 
P\'azm\'any P\'eter s\'et\'any 1/C H-1117 Budapest, Hungary
}{
\href{mailto:patakino@elte.hu}{patakino@elte.hu}
}
%%
%% MORE AUTHORS. For more than two authors, please, use both commands, maybe more than once.
%%

%% Short name of the authors and short title, to be included in heading.
\short{%
%% Write here the short name of authors, using commas.
N. Pataki
}{%
%%Write here the short title
C++ Standard Template Library by template specialized containers
}

\begin{abstract}
The C++ Standard Template Library is the flagship example for libraries
based on the generic programming paradigm.  The usage of this library 
is intended to minimize the number of classical C/C++ errors, but  does
not warrant  bug-free programs. Furthermore, many new kinds of errors may
arise from the inaccurate use of the generic programming paradigm, like
dereferencing invalid iterators or misunderstanding remove-like algorithms.

In this paper we present some typical scenarios that may cause runtime or 
 portability problems. We emit warnings and errors while these
risky constructs are used. We also present a general approach to emit ``customized''
warnings. We support the so-called ``believe-me marks'' to disable warnings.
We present another typical usage of our technique, when classes
become deprecated during the software lifecycle.
\end{abstract}

\section{Introduction}
\label{intro}
The \emph{C++ Standard Template Library} (STL) was developed by \emph{generic
programming} approach \cite{austern:stl}. In this way containers are defined as
class templates and many algorithms can be implemented as function templates.
Furthermore, algorithms are implemented in a container-independent way, so one
can use them with different containers \cite{stroustrup:cpp}.
C++ STL is widely-used because it is a very handy, standard
C++ library that contains beneficial containers (like list,
vector, map, etc.) and a large number of algorithms (like sort,
find, count, etc.) among other utilities \cite{czarnecki:generative}.

The STL was designed to be extensible \cite{musser:generic}. 
We can add new containers that can work together with the existing algorithms. 
On the other hand, we can extend the set of algorithms with a new one that can
work together with the existing containers. Iterators bridge the gap between 
containers and algorithms \cite{becker:iterators}. The expression
problem \cite{torgersen:expr} is solved  with this approach. STL also includes
adaptor types which transform standard elements of the library for
a different functionality \cite{alexandrescu:modern}.

However, the usage of C++ STL does not guarantee bugless or error-free code
\cite{devai:tool}. Contrarily, incorrect application of the library may
introduce new kinds of problems \cite{stlmetric}.

One of the problems is that the error diagnostics are usually complex,
and very hard to figure out the root cause of a program error
\cite{zolman:message,zolyomi:introspection}. Violating requirement of
special preconditions (e.g. sorted ranges) is not tested, but results in
runtime bugs \cite{pataki:overhead}. A different kind of stickler is that
if we have an iterator object that pointed to an element in a container, but the
element is erased or the container's memory allocation has been changed, then
the iterator becomes \emph{invalid} \cite{pataki:ranges}. Further reference of
using invalid iterators causes undefined behaviour \cite{pataki:safestl}.

Another common mistake is related to removing algorithms. The algorithms are
container-independent, hence they do not know how to erase elements
from a container, just relocate them to a specific part of the container,
and we need to invoke a specific erase member function to remove the
elements phisically. Since, for example the \texttt{remove} algorithm
does not actually remove any element from a container \cite{meyers:STL}.

Some of the properties are checked at compilation time \cite{gregor:concepts}.
For example, the code does not compile if one uses sort algorithm with the
standard list container, because the list's iterators do not
offer random accessibility \cite{jarvi:specialization}. Other
properties are checked at runtime \cite{pirkelbauer:runtime}, like
the standard vector container offers an \texttt{at} method
which tests if the index is valid and it raises an exception
otherwise \cite{pataki:soundness}.

Unfortunately, there are still a large number of properties that
are tested neither at compilation-time nor at run-time. The observance
of these properties is in the charge of the programmers \cite{devai:stl}.
On the other hand, type systems can provide a high degree of safety at low
operational costs. As part of the compiler, they discover many semantic
errors very efficiently.

Associative containers (e.g. \texttt{multiset}) use functors exclusively to
keep their elements sorted. Algorithms for sorting (e.g. \texttt{stable\_sort})
and searching in ordered ranges (e.g. \texttt{lower\_bound}) are typically used
with functors because of efficiency. These containers and algorithms need
\emph{strict weak ordering}. Containers become inconsistent if used
functors do not meet the requirement of strict weak ordering \cite{pataki:functors}.

%Name of some algorithms are can be found as a member function in certain containers.
Certain containers have member functions with the same names as STL algorithms. This
phenomenon has many different reasons, for instance efficiency, safety or avoidance
of compilation errors. For example, as mentioned before list's iterators cannot
be passed to \texttt{sort} algorithm, hence code with this mistake  cannot be
compiled \cite{lupin:modularization}. To overcome this problem list has a member
function called \texttt{sort}. In these cases, although the code compiles, the
member function calls are preferable to the usage of generic algorithms.

Whereas C++ STL is pre-eminent in a sequential realm, it is not aware of
multicore environment \cite{austern:range}. For example, the Cilk++ language
aims at multicore programming. This language extends C++ with new keywords
and one can write programs for multicore architectures easily. Although
the language does not contain an efficient multicore library, just
the C++ STL only which is an efficiency bottleneck in multicore environment.
We develop a new STL implementation for Cilk++ to cope with the challenges
of multicore architectures\cite{lupin:multicore}. This new implementation can
be safer solution, too. Hence, our safety extensions will be included in the
new implementation. However, the advised techniques presented in this paper
concern to the original C++ STL, too.

In this paper we argue for an approach that generates warnings or errors
when a template container is instantiated with improper parameters. These
instantiations mean erroneous, unportable code or other weird compilation effects.
A general technique is presented to express custom warnings at compilation time.
Our technique is able to indicate the usage of deprecated classes.

This paper is organized as follows. In Section \ref{warnings} we present an approach
to generate ``customized'' warnings at compilation time. After, in Section
\ref{vectorbool} we describe the specialized vector container which contains boolean
values. We show why this container is problematic, and argue for warnings when it is
in use. We explain the forbidden \emph{containers of auto pointers} and present
an approach to disable their usage by template specializations. In Section \ref{marks}
the so-called \emph{believe-me marks} are introduced. Finally, this paper concludes
in Section \ref{conc}.

\section{Generation of warnings}
\label{warnings}

Compilers cannot emit warnings based on the erroneous usage of the library.
STLlint is the flagship example for external software that is able to emit
warnings when the STL is used in an incorrect way \cite{gregor:stllint}.
We do not want to modify the compilers, so we have to enforce the compiler
to indicate these kinds of potential problems. However, \texttt{static\_assert}
as a new keyword is introduced in C++0x to emit compilation errors based on 
conditions, but no similar construct is designed for warnings.

\begin{verbatim}
template <class T>
inline void warning( T t ) { }

struct VECTOR_BOOL_IS_IN_USE { };

// ...

warning( VECTOR_BOOL_IS_IN_USE() );
\end{verbatim}

When the \texttt{warning} function is called, a dummy object is passed. This dummy object
is not used inside the function template, hence this is an unused parameter. Compilers
emit warning to indicate unused parameters. Compilation of \texttt{warning} function
template results in warning messages, when it is referred and
instantiated \cite{pataki:funcframework}. No warning message is shown
if it is not referred. In the warning message the template argument is
referred. New dummy type has to be written for every new kind of warning.

Different compilers emit this warning in different ways. For instance, Visual Studio emits
the following message:

\begin{verbatim}
warning C4100: 't' : unreferenced formal parameter
...
see reference to function template instantiation 'void
warning<VECTOR_BOOL_IS_IN_USE>(T)'
being compiled

        with
        [
            T=VECTOR_BOOL_IS_IN_USE
        ]
\end{verbatim}

And g++ emits the following message:

\begin{verbatim}
In instantiation of 'void warning(T)
      [with T = VECTOR_BOOL_IS_IN_USE]':
... instantiated from here
... warning: unused parameter 't'
\end{verbatim}

Unfortunately, implementation details of warnings may differ, thus there is no
universal solution to generate custom warnings.

This approach of warning generation has no runtime overhead
inasmuch as the compiler optimizes the empty function body. On the other
hand---as the previous examples show---the message refers to the
warning of unused parameter, incidentally the identifier of the template
argument type is appeared in the message.

\section{The weirdest vector}
\label{vectorbool}

In this section we present the basic idea behind the specialized \texttt{vector<bool>}
container. We present the pros and cons of this weird type. We argue for generate
warnings at compilation-time if a programmer uses \texttt{vector<bool>} because
it is the embodiment of the weird container.

Many programmers think that the \texttt{vector<bool>} is the instantiation of
STL's \texttt{vector} template, but it is not true. On many platforms
\texttt{sizeof( int ) == sizeof( bool )} because of reverse compatibility.
(In the C programming language \texttt{int} type has been used to represent Boolean
values.) Hence, the \texttt{vector<bool>} is a template specialized container
to develop a more advanced, denser implementation for boolean values.
This representation is able to represent 32 boolean values on 4 bytes.

The following code sketch represents the connection between \texttt{vector<bool>}
and \texttt{vector} template:

\begin{verbatim}
template <class T, class Alloc = std::alloc>
class vector
{
  T* p;
  size_t capacity;
  size_t size;
public:
  vector()
  {
    // ...
  }

  void push_back( const T& t )
  {
    // ...
  }
  
  // ...
};

template <class Alloc>
class vector<bool, class Alloc>
{
  // dense representation of vector bool
  // No bool* member
public:
  // public interface is similar to the previous one

  void push_back( const bool& t )
  {
    // ...
  }

  vector()
  {
    //...
  }
};
\end{verbatim}

So, the \texttt{vector<bool>} has a special representation to handle dense boolean
values. It is designed to be effective when someone stores boolean values. But it
has weird behaviour compared to the \texttt{vector} template:

\begin{verbatim}
std::vector<int> a;
a.push_back( 3 );
int* p = &a[0]; 

std::vector<bool> b;
b.push_back( true );
bool* q = &b[0];
\end{verbatim}

The previous code does not compile because of the \texttt{bool* q = \&b[0];} assignment.
However, when the template \texttt{vector} is in use, its counterpart does compile.
It is a contradiction in terms, because this way the \texttt{vector<bool>} cannot
meet the requirements of C++ Standard. Hence, it is not advised to use.
Let us see the background of this compilation issue:

\begin{verbatim}
template <class T, class Alloc = std::alloc>
class vector
{
  T* p;
  //..
public:
  T& operator[]( int idx )
  {
    return p[idx];
  }
  
  const T& operator[]( int idx ) const
  {
    return p[idx];
  }
  // ...
};

template <class Alloc>
class vector<bool, class Alloc>
{
  // dense representation of vector bool
  // No bool* member
public:

  class bool_reference
  {
     // ...
  };

  bool_reference operator[]( int idx )
  {
     // ...
  }
};
\end{verbatim}

Because the \texttt{vector<bool>} does not hold actual bool values it cannot return
\texttt{bool\&}. Hence, a proxy class is developed which actually simulates
\texttt{bool\&}. However, conversions cannot be defined between \emph{pointer to a
\texttt{bool\_reference}} and a \emph{pointer to a \texttt{bool}}.
This behaviour can be much more appalling, when the programmer uses \texttt{vector}
as a base class. Arcane error messages are emitted when the subtype is instantiated
with \texttt{bool}.

Unfortunately, most of STL references hardly mention that \texttt{vector<bool>} is
not the instantiation of template, but a completely different class.
It would be useful if the compiler indicated if the programmer used \texttt{vector<bool>}
container, even intentionally or inadvertently.

Now it is not difficult to emit warning with the presented function.
Fortunately, \texttt{vector<bool>} is still a class template because the
type of its allocator is a template parameter. So, the compilation
warning is emitted only when this template class is instantiated, hence
someone uses it:

\begin{verbatim}
template<class Allocator>
class vector<bool, Allocator>
{
  // ...
public:
  vector()
  {
    warning( VECTOR_BOOL_IS_IN_USE() );
    // ...
  }

  template<class InputIterator>
  vector( InputIterator first, InputIterator last )
  {
    warning( VECTOR_BOOL_IS_IN_USE() );
    // ...
  }

  vector( size_t n, const bool& value = bool() )
  {
    warning( VECTOR_BOOL_IS_IN_USE() );
    // ...
  }

  vector( const vector& rhs)
  {
    warning( VECTOR_BOOL_IS_IN_USE() );
    // ...
  }

};
\end{verbatim}

In Section \ref{warnings} the emitted warning message can be seen.

\section{Containers of auto pointers}
\label{coaps}

In this section the containers of auto pointers are detailed.
We present their motivation and reason why are they problematic.
We present a solution to forbid the usage of these kinds of containers.

Usually, auto pointers (\texttt{std::auto\_ptr} objects) make easier to
manage objects in the heap memory. This class assists in memory management.
The auto pointers deallocate the pointed memory when they are gone out
of scope \cite{stroustrup:cpp}. Hence, they prevent memory leaks:

\begin{verbatim}
void f()
{
  std::auto_ptr<int> p( new int( 5 ) );
  // no memory leak
}
\end{verbatim}

Containers of STL are template classes, so technically they should be 
instantiated with auto pointers and store auto pointers that point to the heap:

\begin{verbatim}
std::vector<std::auto_ptr<int> > v;
v.push_back( new int( 7 ) );
// ...
\end{verbatim}

The previous code snippet seems to be safe. On the other hand, the C++
Standardization Committee forbid the usage of \emph{containers of auto pointers (COAPs)}.
The motivation behind this idea is that the copy of auto pointers is strange:

\begin{verbatim}
std::auto_ptr<int> p( new int( 3 ) );
std::auto_ptr<int> q = p;
// At this point p is null pointer
\end{verbatim}

The copied auto pointer becomes null pointer. Only one auto pointer is able to
point to any object in the heap. This one is responsible for the deallocation.

So, if COAPs are not be forbidden, the following code snippet results in a very
strange behaviour:

\begin{verbatim}
struct Auto_ptr_less
{
  bool operator()( const std::auto_ptr<int>& a, 
                   const std::auto_ptr<int>& b )
  {
    return *a < *b;
  }
};

std::vector<std::auto_ptr<int> > v;
v.push_back( new int( 7 ) );
// ...
std::sort( v.begin(), v.end(), Auto_ptr_less() );
\end{verbatim}

Some of the pointers may become null pointer because of the assignments during
swapping vector's elements when it is necessary. This is the reason why COAPs
are forbidden.

Unfortunately, some of the compilers and STL platforms are still permitting
the usage of COAPs, some of them are not. This inhibits the writing of
portable code \cite{meyers:STL}.

We argue for an extension to emit compilation error if COAPs are in use.
We have to create specializations for auto pointers. The trick that is
we do not write the implementation for the auto pointer specializations.
Thus, these specializations are declared, but are not defined types.
For instance, the vector declaration can be the following:

\begin{verbatim}
template <class T, class Alloc>
class vector< std::auto_ptr<T>, Alloc>;
\end{verbatim}

The instantiation of a COAP results in the hereinafter error message:

\begin{verbatim}
error: aggregate 'std::vector<std::auto_ptr<int>,
  std::allocator<std::auto_ptr<int> > > v'
  has incomplete type and cannot be defined
\end{verbatim}

We have to develop these declarations for all standard containers.
These declarations mean bugless and more portable code.

\section{Believe-me marks}
\label{marks}

Generally, warnings should be eliminated. On the other hand, the usage of
\texttt{vector<bool>} does not necessarily mean a problem. It can be used
safely. However, we cannot disable the generated warning if it is in use.

Believe-me marks \cite{kto:subtype} are used to identify the points in the program
text where the type system cannot obtain if the used construct is risky. For instance,
in the hereinafter example, the user of the library asks the type system to ``believe''
that the programmer is conscious of the specialized vector container. This way we
enforce the user to reason about the parameters of containers.

First, we create a new type which stands for the believe-me mark:

\begin{verbatim}
struct I_KNOW_VECTOR_BOOL { };
\end{verbatim}

After, we extend the vector template container with one new template parameter.
The new template parameter has default parameter value, so it is reverse compatible
with the original container. This parameter has not been taken advantage of, and
has no effect on the implementation:

\begin{verbatim}
template <class T, class Alloc = std::alloc, class Info = int>
class vector { };
\end{verbatim}

Let us consider, that the original implementation of \texttt{vector<bool>}
which does not generate warning has been removed to a new template class:

\begin{verbatim}
template <class Alloc>
class __VectorBool
{
  // original implementation of vector<bool>
};
\end{verbatim}

The new template parameter has effect on the \texttt{vector<bool>} specialization:

\begin{verbatim}
template <class Alloc>
class vector<bool, I_KNOW_VECTOR_BOOL, Alloc>: 
  public __VectorBool<Alloc>
{ };

template <class Alloc>
class vector<bool, Alloc, I_KNOW_VECTOR_BOOL>: 
  public __VectorBool<Alloc> { };

template <class Alloc, class Info>
class vector<bool, Alloc, Info>: 
  public __VectorBool<Alloc>
{
public:
  vector(): __VectorBool<Alloc>()
  {
    warning( VECTOR_BOOL_IS_IN_USE() );
  }

  template<class InputIterator>
  vector( InputIterator first, InputIterator last ): 
    __VectorBool<Alloc>( first, last )
  {
    warning( VECTOR_BOOL_IS_IN_USE() );
  }

  vector( size_t n, const bool& value = bool() ):
    __VectorBool<Alloc>( n, value )
  {
    warning( VECTOR_BOOL_IS_IN_USE() );
  }

  vector( const vector& rhs): __VectorBool<Alloc>( rhs )
  {
    warning( VECTOR_BOOL_IS_IN_USE() );
  }
};
\end{verbatim}

In this case no compilation warning is emitted if the
last added template parameter is \texttt{I\_KNOW\_VECTOR\_BOOL},
otherwise the mentioned warning can be seen during compilation.

\vspace*{-1em}
\section{Deprecated classes}
\label{deprecated}

In section \ref{vectorbool} we generated warnings when a template-specialized class
was used. A similar idea can be mentioned. It would be useful to generate warnings
when the usage of classes becomes unsupported.

A common idea during a software lifecycle is, that some of the classes are not deleted
from the project, but their usage is not advised. These classes are called
\emph{deprecated}. Deprecated annotation can be added to classes in Java.
Instantiation of deprecated classes results in compilation
warnings \cite{juhasz:teachingjava}. However, no similar technique is used in C++.

First, we create some utility classes for warning generation:
\begin{verbatim}
struct DeprecatedClass { };

template <class DEPRECATED>
struct Deprecated
{
  Deprecated()
  {
    warning( DeprecatedClass() );
  }
};
\end{verbatim}

The role of the template parameter in \texttt{Deprecated} struct is to pass the
identifier of deprecated class to the emitted warning.

Now, let us consider that the following class becomes deprecated during software
lifecycle:

\begin{verbatim}
class Foo
{
  // ...
public:
  Foo( int a, int b)
  {
    // ...
  } 
};
\end{verbatim}

The user has to add one more base class to the deprecated class. This does not
mean limitation because the C++ programming language supports multiple inheritance.
For example:

\begin{verbatim}
class Foo: public Deprecated<Foo>
{
  // very same...
};
\end{verbatim}

The following warning is received from the compiler:

\begin{verbatim}
In instantiation of 'void warning(T)
  [with T = DeprecatedClass]':
...   instantiated from
  'Deprecated<DEPRECATED>::Deprecated()
  [with DEPRECATED = Foo]'
...   instantiated from here
... warning: unused parameter 't'
\end{verbatim}

However, this message is received irrespectively of its usage. If the usage
is important, the deprecated class or a called method or constructor must be
a template. This transformation cannot be executed automatically with the
respect of client code. Our future work is to overcome this situation.

We do not advise to make believe-me marks for the deprecated classes inasmuch as
always exists a better approach to use.

\section{Conclusions}
\label{conc}

C++ STL is the most widely-used library based on the generic programming paradigm. It
is efficient and convenient, but the incorrect usage of the library results in weird
or undefined behaviour.

In this paper we argue for some extensions to make the STL itself safer.
Not supported or not advised instantiations result in compilation warnings
and errors to prevent unportable or defective code.

We present an effective approach to generate custom warnings. Believe-me marks are
also written to disable warning messages. With our technique classes can be
marked deprecated, too.

\section*{Acknowledgements}

The Project is supported by the European Union and co-financed by
the European Social Fund (grant agreement no. T\'AMOP
4.2.1/B-09/1/KMR-2010-0003).

%% All references must be cited in the text of your paper. In references, after a red arrow will appear  
%% the hyperlink backreferences to the pages where they have been cited. 

%% For the usual abbreviation of theoretical journals you can use the address: http://www.ams.org/msnhtml/serials.pdf

\bigskip
\rightline{\emph{Received: March 16,  2011 {\tiny \raisebox{2pt}{$\bullet$\!}} Revised: July 20, 2011}} %% to be completed by the editor

\end{document}